\documentclass[aps,prd,12pt,preprint,preprintnumbers,superscriptaddress,amsmath,amssymb,floatfix,nofootinbib]{revtex4-1}
\usepackage{amsmath,latexsym,amssymb,graphicx,hyperref,color,enumerate}
\usepackage{epsf, array, color}
\usepackage{graphicx}
\usepackage{subfigure}
\usepackage{bbold}
\usepackage{amsmath}
\usepackage{mathtools}
\usepackage{dcolumn}% Align table columns on decimal point
\usepackage{footnote}

\newcommand{\bea}{\begin{eqnarray}}
\newcommand{\eea}{\end{eqnarray}}

\def\beq{\begin{equation}}
\def\eeq{\end{equation}}

\hypersetup{colorlinks,citecolor= nicegreen,linkcolor= nicered}
\usepackage{color}
\definecolor{nicered}{rgb}{0.7,0.1,0.1}
\definecolor{nicegreen}{rgb}{0.1,0.5,0.1}

\begin{document}

\title{Higgs CP Violation from Vectorlike Quarks}
%\title{Aspects of CP violation from Vectorlike Quarks}
\author{Chien-Yi Chen}
\author{S. Dawson}
\affiliation{
Department of Physics, Brookhaven National Laboratory, Upton, N.Y., 11973,  U.S.A.}
\author{Yue Zhang}
\affiliation{Walter Burke Institute for Theoretical Physics, \\
California Institute of Technology, Pasadena, CA 91125}

%\author{Chien-Yi~Chen$^{\, a}$, S.~Dawson$^{\, a}$ and Yue~Zhang$^{\, b}$ }
%
%\affiliation{
%\vspace*{.5cm}
%$^a$Department of Physics,
%  Brookhaven National Laboratory, Upton, N.Y., 11973,  U.S.A.\\
% \mbox{$^b$ Walter Burke Institute for Theoretical Physics, \\
%California Institute of Technology,\\ 
%Pasadena, CA 91125}
% \vspace*{1cm}}

\date{\today}

\preprint{CALT-TH-2015-042}

\begin{abstract}
We explore CP violating aspects  in the Higgs sector of models where new vectorlike quarks carry Yukawa couplings mainly to the third generation quarks of the Standard Model. We point out that in the simplest model,  Higgs CP violating interactions only exist in the $hWW$ channel. At low energy, we find that rare $B$ decays can place similarly strong constraints as those from
 electric dipole moments on the source of CP violation. These observations offer a new handle to discriminate from  other Higgs CP violating scenarios such as scalar sector extensions of the Standard Model, and imply 
an interesting future interplay among limits from different  experiments.
\end{abstract}

\maketitle
\section{Introduction}
\label{introsec}
Now that a Higgs boson has been observed with properties similar to those predicted by the Standard
Model (SM), the next critical task is a program of precision measurements of its properties.  Studies of the
Higgs mass, coupling strengths, and production and decay channels are well advanced.  
However, less attention
has been paid to the possibility of observing CP violation in the Higgs sector and
the purpose of the present work is to explore  such a possibility.

If the 125 GeV boson is measured to be a mixture of CP even and odd states, it immediately indicates that
there must be new physics not far above the electroweak scale.
One of the most straightforward ways is to extend the scalar sector of the SM, and
the simplest case is the complex version of the 2 Higgs doublet model 
(C2HDM)~\cite{Fontes:2015gxa,Fontes:2015xva,Fontes:2015mea,Lavoura:1994fv,Grzadkowski:2014ada,Gunion:2005ja}.
The presence of CP violation leads to observable changes in Higgs production and decay rates, as well as contributions
to low energy observables such as electric dipole moments~\cite{Inoue:2014nva}.  
It has also been pointed out that some of the heavy scalar decay channels could be sensitive to CP violation~\cite{Chen:2015gaa}, and could be probed at the LHC and future colliders.
The complementarity of LHC and low energy measurements for constraining Higgs CP violation has been 
explored in Refs.~\cite{Shu:2013uua,Inoue:2014nva,Chen:2015gaa,Cheung:2014oaa}.

Alternatively, it is also possible to extend the fermion sector of the SM while keeping the scalar sector minimal.
The simplest case is to introduce a chiral fourth generation or mirror family, but they are strongly disfavored after the discovery of the Higgs boson, since they would lead to a large
enhancement in the Higgs production rate. The next simplest extension is to introduce vectorlike quarks (VLQs). 
Vectorlike fermions are defined as having the same gauge quantum numbers for both left- and right-handed fermion pairs, and  thus they do not generate chiral anomalies and their effects decouple in Higgs physics. They are the ingredients of beyond the SM frameworks like the little Higgs~\cite{ArkaniHamed:2002qx} or composite Higgs models~\cite{Kaplan:1983sm, Agashe:2004rs}, and theories of extra dimensions~\cite{ArkaniHamed:1998rs,Randall:1999ee,Appelquist:2000nn} or extended supersymmetry
~\cite{Fayet:1984wm,delAguila:1984qs}.
VLQs have also been discussed recently in light of modified Higgs couplings such as that to two
photons~\cite{Voloshin:2012tv,Fan:2013qn}.~\footnote{There are also recent studies which extend both scalar and fermion sectors of the SM~\cite{McKeen:2012av, Chao:2015uoa, Gopalakrishna:2015wwa}.} 
The current LHC lower limit on the masses of VLQs from direct searches is around 800\,GeV~\cite{Aad:2015kqa} regardless of the decay modes.
The indirect effects of VLQs in  electroweak precision measurements and flavor physics have also been extensively explored in the literature~\cite{Dawson:2012di, Aguilar-Saavedra:2013qpa, Fajfer:2013wca, Ellis:2014dza,Efrati:2015eaa}.

%  Vectorlike fermions are 
%defined as having the same gauge quantum numbers for both left- and right-handed projections. 
%%
%Therefore, the introduction of vctorlike fermions will not give rise to chiral anomaly. 
%Vectorlike fermions can naturally arise in many New Physics models,
%such as the composite Higgs models~\cite{Agashe:2004rs,Marzocca:2012zn}, the composite vectorlike model~\cite{Dobrescu:2014zla}, and
%models of extra dimensions with fermions in the bulk~\cite{Cheng:2010pt,Ponton:2012bi}, where the zeroth mode of KK particles are chiral, 
%which correspond to the SM particles, while the higher excited states are vectorlike fermions.
%Also, in the little Higgs models a vectorlike quark is introduced to 
%cancel the UV divergence of the Higgs self enery resulting from the top quark loop~\cite{ArkaniHamed:2001nc,ArkaniHamed:2002qx}.
%

In this work, we consider the CP violating aspects of the VLQ models, which have been less studied.
Our goal is to examine their impact on the CP nature and interactions of the Higgs boson, as well as the low energy constraints\cite{Cirigliano:2013lpa}. For simplicity, we focus on the simple cases where VLQs are in a single representation of $SU(3)_c\times SU(2)_L\times U(1)_Y$.
We consider only fermion representations that can have new Yukawa couplings
with the SM quarks and Higgs doublet.
These new Yukawa couplings could provide
a new source of CP violation at the weak scale. Under these assumptions, we find that only the case where the VLQ lies 
in the $(3,2,1/3)$ representation can generate  significant CP violation in Higgs physics.

We study  CP violating phenomenology in the doublet VLQ model using an effective  theory language where the heavy VLQs have been integrated out. Then the CP violating effects manifest themselves through a new right-handed charged-current interaction mediated by the $W$-boson. We clarify the source of CP violation in this model in Section \ref{model}.
In Section \ref{higgs}, we calculate the loop induced CP violating Higgs interactions with gauge bosons.
Interestingly, we find that CP violation only exists in the $hWW$ coupling, but not in the $hZZ$\cite{Christensen:2010pf}, 
$h\gamma \gamma$, or $hZ\gamma$\cite{Choi:2012yg}
 ones. In Section \ref{const}, we explore the current constraints on this coupling from low energy measurements, including EDMs and $B$ physics, 
and comment on the future prospects.

\section{CP Violation from Vectorlike Quarks}
\label{model}
The vectorlike quark representations that allow new Yukawa couplings with SM quarks are summarized in Table~\ref{table1}\footnote{We find that introducing any number of singlet $T_{L,R}$ and/or $B_{L,R}$ fields does not give rise to new CP violating phases since any potential new phases can always be rotated away by a field redefinition.}.

\begin{table}[h]
\centering
\begin{tabular}{|c|c|c|}
\hline
VLQ models & Representation & CP violation \\
\hline
$T_{L,R}$ & $(3,1,4/3)$ & no\\
\hline
$B_{L,R}$ & $(3,1,-2/3)$ & no \\
\hline
$(T, B)_{L,R}$ & $(3,2,1/3)$ & \textcolor[rgb]{1,0,0}{yes} \\
\hline
$(X, T)_{L,R}$ & $(3,2,7/3)$ & no \\
\hline
$(B, Y)_{L,R}$ & $(3,2,-5/3)$ & no \\
\hline
$(X, T, B)_{L,R}$ & $(3,3,4/3)$ & no \\
\hline
$(T, B, Y)_{L,R}$ & $(3,3,-2/3)$ & no \\
\hline
\end{tabular}
\caption{Models of vectorlike quarks and their representations under $SU(3)_c\times SU(2)_L
\times U(1)_Y$, together with the possibilities of introducing new physical CP violating
 phases.}\label{table1}
\end{table}

The key point we want to make is that only the doublet $(T, B)$ model can offer non-zero (unsuppressed) CP violation for the Higgs boson, through the Yukawa coupling to third generation quarks\footnote{We assume only one representation of VLQs is
present.}.
To see this, we write the Yukawa sector of this model,
\begin{eqnarray}\label{eq:Ldoublet}
\mathcal{L}_Y &=& y_t \bar Q_{3L} \tilde H t_R + y_b \bar Q_{3L} H b_R + M \bar Q'_L Q'_R  + M' \bar Q_{3L} Q'_R+ \lambda_t \bar Q'_L \tilde H t_R + \lambda_b \bar Q'_L H b_R + {\rm h.c.} \,,
\label{Yukdef}
\end{eqnarray}
where $H=(\phi^+, \phi_0)^T$ is the SM Higgs doublet, $\tilde H = i\sigma_2 H^*$, $Q_{3L}^T = (t_L, b_L)$ is the third generation left-handed quark doublet and $Q'^T_{L,R}=(T, B)_{L,R}$ are the vectorlike quark doublets.

In general, the SM gauge invariance permits us to generalize the fields above, ($Q_{3L}, t_R, b_R$), to linear combinations of all  three generations.
In our study, we assume the fields of Eq.~(\ref{Yukdef}) are dominantly composed of third 
generation fermions, because they have the largest Yukawa couplings  and thus have the strongest impact on CP violation in Higgs physics, which is the motivation of this work. 
To be concrete, we can take advantage of the hierarchical structure of the CKM matrix, and define Eq.~(\ref{eq:Ldoublet}) in the basis where the SM $3\times3$ blocks of the up- and down-type Yukawa matrices are close to diagonal up to CKM-like rotations\footnote{Alternatively, we
could work in the basis where the $3\times3$ block of down quark Yukawa couplings is already diagonal. With this assumption, the down quark sector is free from new contributions to flavor violating processes.}. This helps to suppress the mixing between heavy VLQs and the first two generation quarks and minimize the low energy flavor changing effects in the spirit of next-to-minimal flavor violation~\cite{Agashe:2005hk}.

%To be concrete, we can define Eq.~(\ref{eq:Ldoublet}) in the basis where the SM $3\times3$ block of the down-type quark Yukawa matrix is diagonal, 
%and the counterpart in the up-type Yukawa (approximately) differs from being diagonal up to a CKM(-like) rotation from left and right.
%In this basis, the mixing between heavy VLQs and first two generation quarks
%This helps to suppress the flavor-changing effects in light meson d
%Throughout the discussion, we will neglect the mixings of first two generation quarks and the new VLQs.
%The CKM matrix being close to diagonal allows one to choose a basis where these mixings are suppressed by powers of $\lambda$, $\lambda=0.22$ is the Wolfenstein parameter.
%The light quark contributions to new Higgs interactions is always subdominant.
%Finally, the diagonalization of the $3\times3$ up-type quark Yukawa matrix should give the usual CKM mixing. Moreover, because $t_R$ gets gauge interaction in effective theory (given by Eq.~(\ref{eq:Leff})),
%its mixings with light quarks $u_R, c_R$ (which are pure $SU(2)_L$ singlets) now become physical. The light quarks have little effect on Higgs CP violation but their mixings with $t_R$ could be constrained by flavor physics like $D-\bar D$ mixing.
%For simplicity we further assume they are small.

From now on, we will focus on the mixing between VLQs and the third generation quarks. 
Since $Q_{3L}$ and $Q'_L$ have the same quantum numbers, one can always redefine fields and set the parameter $M'=0$.
After electroweak symmetry breaking, the quark mass matrices take the form,
\begin{eqnarray}\label{eq:mass}
\mathcal{L}_m &=& (\bar t_L,\ \bar T_L)
\begin{pmatrix}
\frac{y_t v}{\sqrt{2}} & 0 \\
\frac{\lambda_t v}{\sqrt{2}} & M
\end{pmatrix}
\begin{pmatrix}
t_R \\
T_R
\end{pmatrix} + (\bar b_L,\ \bar B_L)
\begin{pmatrix}
\frac{y_b v}{\sqrt{2}} & 0 \\
\frac{\lambda_b v}{\sqrt{2}} & M
\end{pmatrix}
\begin{pmatrix}
b_R \\
B_R
\end{pmatrix} \, .
\end{eqnarray}
In general the parameters are all complex, and one can remove unphysical phases by redefining the phases of the fields. Under the gauge invariant transformations, $Q'_{R} \to Q'_{R} e^{i \alpha}$, $Q'_{L} \to Q'_{L} e^{i \beta}$, $Q_{3L} \to Q_{3L} e^{i \gamma}$, $t_{R} \to t_{R} e^{i \delta}$, $b_{R} \to b_{R} e^{i \sigma}$, the parameters change to
\begin{eqnarray}
&&y_t \to y_t e^{i(\delta - \gamma)}\, , \ \ \ \lambda_t \to \lambda_t e^{i(\delta - \beta)}\, , \nonumber \\
&&y_b \to y_b e^{i(\sigma - \gamma)}\, , \ \ \ \lambda_b \to \lambda_b e^{i(\sigma - \beta)}\, , \nonumber \\
&&M\to M e^{i(\alpha- \beta)}\, .
\end{eqnarray}
Clearly, what is invariant is the combination of phases of the parameters $\arg(y_b) + \arg(\lambda_t) - \arg(y_t) - \arg(\lambda_b)$, or the quantity,
\begin{eqnarray}\label{eq:CPV}
{\rm Im}(y_b \lambda_t y_t^* \lambda_b^*) \equiv |y_t y_b \lambda_b \lambda_t| e^{i \theta} \, .
\end{eqnarray}
This is the only new source of CP violation in this model that can enter into physical processes. 
It is convenient to use the above rephasing freedom to rotate the phase $\theta$ into $\lambda_t$ and make the other parameters real.
In this case, from the quark mass terms, $y_q (v+h) \bar q q/\sqrt{2}$, we can first assign the Higgs boson to be CP even.
Then any coupling between $h$ and CP odd operators induced by Eq.~(\ref{eq:CPV}) 
indicates CP is violated in the Higgs sector.

We  diagonalize the mass matrices in Eq.~(\ref{eq:mass}), and obtain the mass eigenstates
\begin{eqnarray}
\hat t_R &=& \cos \theta_R^t t_R + \sin\theta_R^t e^{-i\theta} T_R\, , \nonumber \\
\hat b_R &=& \cos\theta_R^b b_R + \sin\theta_R^b B_R\, , 
%\nonumber \\
%\hat t_L &=& \cos\theta_L^ti t_L + \sin\theta_L^t e^{-i\theta} T_L\, , \nonumber \\
%\hat b_L &=& \cos\theta_L^b b_L + \sin\theta_L^b B_L\, , 
\end{eqnarray}
where $\hat T_{R}, \hat B_{R}$ are orthogonal to $\hat t_R$, $\hat b_R$, respectively. 
There are similar mixings among the left-handed fields, parametrized by angles $\theta_L^t$ and $\theta_L^b$. 
The mixing angles among the right-handed quarks satisfy, 
\begin{eqnarray}
\tan2\theta_R^i &=& \frac{\sqrt2 M \lambda_i v}{M^2-(\lambda_i^2+y_i^2)v^2/2}
\simeq \frac{\sqrt{2} \lambda_i v}{M} \,, \ \ \  (i=t, b)\,, \ \
%\nonumber \\
%\tan2\theta_L^i &=& \frac{y_i \lambda_i v^2}{M^2+(\lambda_i^2-y_i^2)v^2/2}
%\simeq \frac{y_i \lambda_i v^2}{M^2}\,, 
\end{eqnarray}
and the last step keeps only the leading term in the $v/M$ expansion.
The angle $\theta_R^b$ denotes the mixing between the  $SU(2)_L$ singlet $b_R$ and doublet $B_R$, and is constrained by electroweak precision measurements such as $Z\to b\bar b$~\cite{Bamert:1996px,Freitas:2014hra,Dawson:2012di}.
The mixings among left-handed quark fields only appear at order $(v/M)^2$ and are much smaller~\cite{Dawson:2012di, Aguilar-Saavedra:2013qpa}.

The lower bound on the VLQ mass scale is around  800\,GeV from
direct searches at the LHC~\cite{Aad:2015kqa}, which suggests that we can integrate them out when studying Higgs physics.
Since  $T_R, B_R$ lie in an $SU(2)_L$ doublet, integrating out the heavy vectorlike quarks yields an anomalous $Wtb$ interaction, 
\begin{equation}
\mathcal{L}_{eff} =a_R \left(\frac{g}{\sqrt2} \right) \bar{\hat{t}}_R \gamma^\mu \hat b_R W^+_\mu\, ,
\end{equation}
where 
\begin{eqnarray}\label{eq:aR}
a_R=\sin\theta_R^t \sin\theta_R^b e^{i\theta} \simeq \frac{|\lambda_b \lambda_t| v^2}{2M^2} e^{i\theta}\,.
\end{eqnarray}

As discussed above, in this model CP violation must appear in physical processes through the combination of couplings, ${\rm Im}(y_b \lambda_t y_t^* \lambda_b^*)$. The new right-handed $Wtb$ coupling $a_R$ obtained here is proportional to $\lambda_t \lambda_b^*$. 
Therefore, a physical process that makes the CP violation manifest should involve both left- and right-handed currents in order to  allow mass (Yukawa coupling $y_t, y_b$) insertions. We write the most general renormalizable $Wtb$ coupling as
\begin{eqnarray}\label{eq:Leff}
 {\cal L}_{eff}= {g \over \sqrt{2}} {\bar t} (a_L P_L +a_R P_R) b W_\mu^+ + h.c. \, .
\end{eqnarray}
We neglect the hat symbol for mass eigenstates hereafter.
In the SM, $a_L = V_{tb} \simeq1$ and $a_R =0$. For the vectorlike quark doublet model we consider, $a_R$ is given by Eq.~(\ref{eq:aR}), and the deviation of $a_L$ from 1 occurs only at higher order
in $v/M$.

Following a similar reasoning, we have also examined  other representations of VLQs. Interestingly, none of them can offer an irreducible CP violating phase such as that in Eq.~(\ref{eq:CPV}), {\it i.e.}, under the same assumptions, the Higgs boson is essentially CP even in those models. 
This observation places the VLQ doublet $(T, B)$ model in a unique place in the perspective of CP violation.
In the coming sections, we will explore the $(T, B)$ model as a theory of Higgs CP violation, and study in detail its predictions in phenomenology and the constraints on the parameters of the model.

\section{CP violation in Higgs Boson Interactions }
\label{higgs}
\begin{figure}[t]
\includegraphics[width=0.65\textwidth]{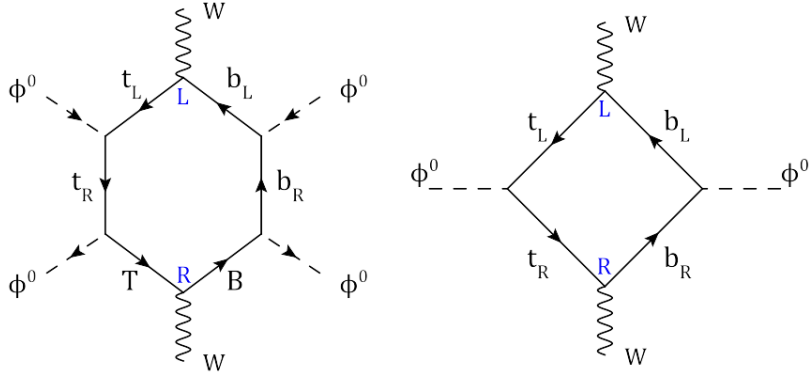}
\caption{Feynman diagrams generating CP violating Higgs couplings. The label ``L (R)'' means a left- (right-) handed current interaction with the $W$-boson. The left diagram is in the full theory, and the right one is in the effective theory when the vectorlike quarks are integrated out. The right-handed current $Wtb$ vertex is derived in Eq.~(\ref{eq:Leff}). For the $hWW$ coupling, 
one of the $\phi^0$ fields is replaced by the electroweak vev, and the other replaced by $h$.}
\label{fig:cpv}
\end{figure}

The Yukawa interactions between vectorlike and SM quarks introduce 
a new source of CP violation. One of the consequences is that the Higgs 
boson will obtain CP violating interactions with the other SM particles.
As discussed in Eq.~(\ref{eq:CPV}), in the model with a single  VLQ doublet, $(T,B)$,
the physical CP violating phase has to appear via
the combination of parameters, ${\rm Im}(y_b \lambda_t y_t^* \lambda_b^*)$.
This means that the diagram giving CP violating interactions to the Higgs boson must 
involve both top and bottom quarks.
As a result, the leading CP violation in this model resides only in the
$h W^+_{\mu\nu} \tilde W^{-\mu\nu}$ operator, generated at loop level as shown in the left diagram of Fig.~\ref{fig:cpv}. At this order, CP violating tree level Higgs-quark or loop level Higgs-$Z$-boson and
Higgs-photon interactions are absent.
The direct probes of CP violating $hWW$ interactions
at the LHC  have been discussed in the $h\to WW$ decay, the $hW$ associated production
channel  and the $WW$ fusion channel~\cite{Bolognesi:2012mm, Delaunay:2013npa}.

%%%%%%%%%%%%%%%%%%%%%%%%%%%%%%%%%
%%%%%%%%%%%%%%%%%%%%%%%%%%%%%%%%

In the heavy fermion limit, the gauge invariant operator generating the Higgs-gauge boson
CP violation starts from dimension 8 in this model,
\begin{equation}
\mathcal{L}_8={C_8\over \Lambda^4} \left[ \epsilon_{ij} H^i (\sigma^a)^j_{\ k} H^k W^a_{\mu\nu} \right]
\left[ \epsilon_{mn} H^m (\sigma^b)^n_{\ l} H^l \tilde W^{b\mu\nu} \right]^*\,,
\end{equation}
$\sigma^{a}$ are the Pauli matrices, $\tilde W_{\mu\nu}=\frac{1}{2}\epsilon_{\mu\nu\alpha\beta} W^{\alpha\beta}$
and $i,j,k,l,m,n=1,2$ are $SU(2)_L$ indices. 
After electroweak symmetry breaking, $H^1\equiv\phi^+=0$ and  $H^2\equiv\phi^0=(v+h)/\sqrt2$ in the unitary gauge.
This projects out the CP violating $h W^+_{\mu\nu} \tilde W^{-\mu\nu}$ interaction
\begin{equation}\label{eq:cutoff}
\mathcal{L}_8\rightarrow {2C_8 v^3\over \Lambda^4}h W_{\mu\nu}^a {\tilde W}^{a *}_{\mu\nu}\equiv a_3^W \frac{h}{v}  W_{\mu\nu}^a {\tilde W}^{a *}_{\mu\nu}\, .
\end{equation}
In the last step, we define the coefficient $a_3^W$ in the same notation as Eq.~(1.15) in the Higgs 
Working Group Snowmass report~\cite{Dawson:2013bba}. 

%%%%%%%%%%%%%%%%%%%%%%%%%%%%%%%%%
\begin{figure}[t]
\includegraphics[width=0.65\textwidth]{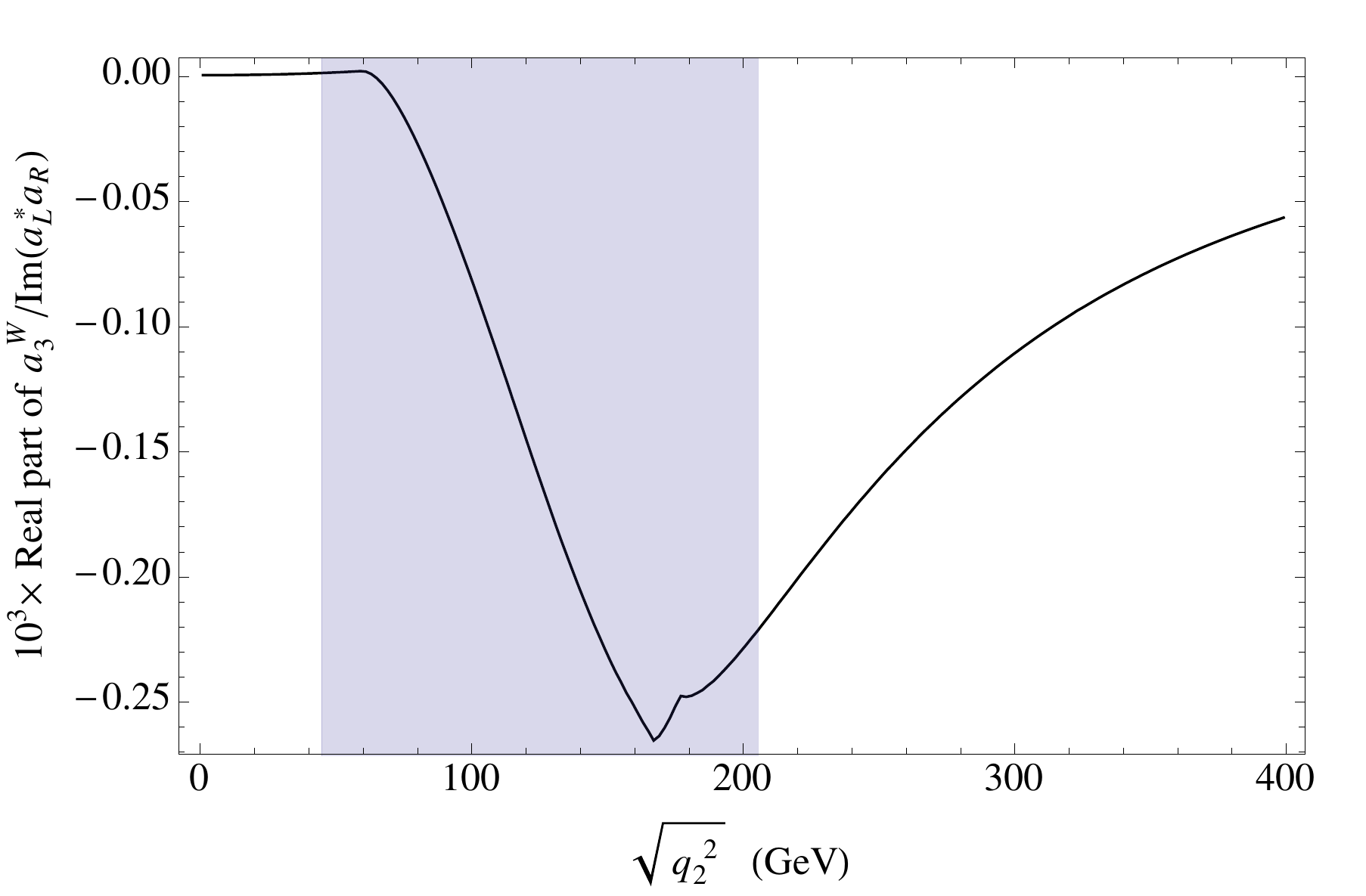}
\caption{Real part of the form factor $a_3^W$ as a function of $\sqrt{q_2^2}$ in units of $10^{-3} \times {\rm Im}(a_L^* a_R)$. For $Wh$ associated production, the kinematics require $\sqrt{q_2^2}>m_h+M_W$, {\it i.e.}, the white region to the right of the shaded region. For $h\to WW^*$ associated production, the kinematics require $0<\sqrt{q_2^2}<m_h-M_W$, {\it i.e.}, the white region to the left of the shaded region.}
\label{formf}
\end{figure}
%%%%%%%%%%%%%%%%%%%%%%%%%%%%%%%%

In reality, because the top quark mass is comparable to the center-of-mass energy $\sqrt{s}$ in the Higgs production and decay processes, the coefficient $a_3^W$ becomes a form factor.
We first calculate the form factor for CP violating $W(q_1)h(p)$ associated production\cite{ Godbole:2014cfa,Godbole:2013saa}
via an off-shell $W^*(q_2)$. In this case, the momenta satisfy $q_1^2=M_W^2$, $p^2=m_h^2$ and $s=q_2^2\geq(m_h+M_W)^2$.
The leading contribution to the form factor $a_3^W$ can be conveniently calculated
in the effective theory when the vectorlike quarks are integrated out, as
shown on the righthand side of Fig.~\ref{fig:cpv}.  
We find
\begin{eqnarray}
a_3^W(\sqrt{q_2^2}) &\simeq&
\frac{\sqrt{2}N_c G_F m_t m_b}{4\pi^2} {\rm Im}(a_L^* a_R) \int_0^1 dx \int_0^{1-x} dy (1-x-y) \left[ \frac{1}{\Delta_t} - \frac{1}{\Delta_b} \right]\,,
\end{eqnarray}
where the denominators are 
\begin{eqnarray}
&&\Delta_t = (x+y)z_t^2 - xy z_h^2 + (x+y-1)(x z^2_1 + y z^2_2 - z^2_b)\,, \nonumber \\
&&\Delta_b = (x+y)z_b^2 - xy z_h^2 + (x+y-1)(x z^2_1 + y z^2_2 - z^2_t)\,,
\end{eqnarray}
and $q_i^2$ are the off-shell momenta of the $W$ gauge bosons, $z_a=m_a/M_W$, $(a=t,h,b)$, $z_1^2=q_1^2/M_W^2=1$,  and $z_2^2=q_2^2/M^2_W=s/M_W^2$.
The $1/\Delta_{t,b}$ terms correspond to the diagrams where the Higgs field is attached to the top (bottom) quark propagators.
There is a minus sign between the two pieces in the integrand of the Feynman parameter integral. 
From the analysis of~\cite{Delaunay:2013npa}, only the real part of the form factor $a_3^W$ contributes to the final
CP violation observable, {\it i.e.}, a phase shift in azimuthal angle.
In Fig.~\ref{formf}, we plot the real part of $a_3^W$ as a function of $\sqrt{q_2^2}$. For $Wh$ associated production, the kinematics require $\sqrt{q_2^2}>m_h+M_W$, {\it i.e.}, the physical region is the white region to the right of the shaded region in the plot.

We next examine the form factor in the decay process $h(p)\to W(q_1)W^*(q_2)$.
In this case, the momenta satisfy $p^2=m_h^2$, $q_1^2=M_W^2$, $0\leq q_2^2\leq (m_h-M_W)^2$.
The integral of the $1/\Delta_t$ term is real. On the other hand, the integral of the $1/\Delta_b$ term has an imaginary (absorptive) part, which is due to a pole in $y$ corresponding to the on-shell cut of the $b\bar b$ propagators. Numerically, we find the integral over $1/\Delta_t$ and dispersive part of the $1/\Delta_b$ integral almost cancel each other.
The integral is dominated by the absorptive part of the $1/\Delta_b$ integral, which we find to be of order 1 for all values of $q_2^2$.  Physically, it indicates that  the CP violating effect in the $h\to WW^*$ decay is dominated by  processes where the Higgs boson first decays to $b\bar b$ and then the $b\bar b$ re-scatter into $WW^*$~\footnote{In general a cut is not necessary for CP violation to occur because the final state $W^+W^-$ is already an eigenstate under CP. The cancelation between the $1/\Delta_t$ and dispersive part of the $1/\Delta_b$ integrals seems accidental.}.

The coefficient $a_3^W$ calculated above is proportional to the quantity ${\rm Im}(a_L^*a_R)$ and is of the order $(v/M)^2$.
There is another contribution obtained by changing the right-handed current to a left-handed one in the heavy quark vertex (left diagram of Fig.~\ref{fig:cpv}), and we have checked that
this contribution is  ${\cal{O}}(v/M)^4$ and is subdominant.

Using the central values of masses and constants from the PDG~\cite{Agashe:2014kda}, we find in both processes the coefficient for the CP violating $hWW$ interaction is
\begin{eqnarray}\label{eq:a3w}
a_3^W \simeq 10^{-3}\times {\rm Im}(a_L^* a_R) \,.
\end{eqnarray}
The first factor contains the usual loop factor and the bottom quark Yukawa coupling. The CP violating parameter ${\rm Im}(a_L^* a_R)$ depends on the model parameters $\lambda_b, \lambda_t$. The goal of the next section is to explore the current and future experimental constraints (sensitivities) to ${\rm Im}(a_L^* a_R)$.

\section{Constraints}
\label{const}

In this section, we explore phenomenological constraints on the parameter ${\rm Im}(a_L^* a_R)$ relevant for the CP violating $hWW$ coupling.
We find the most relevant limits come from the electric dipole moments and the rate and CP asymmetry of the  rare $B$ decay $b\to s\gamma$.

\subsection{Electric Dipole Moments}

\begin{figure}[t]
\centering
\includegraphics[width=0.6\textwidth]{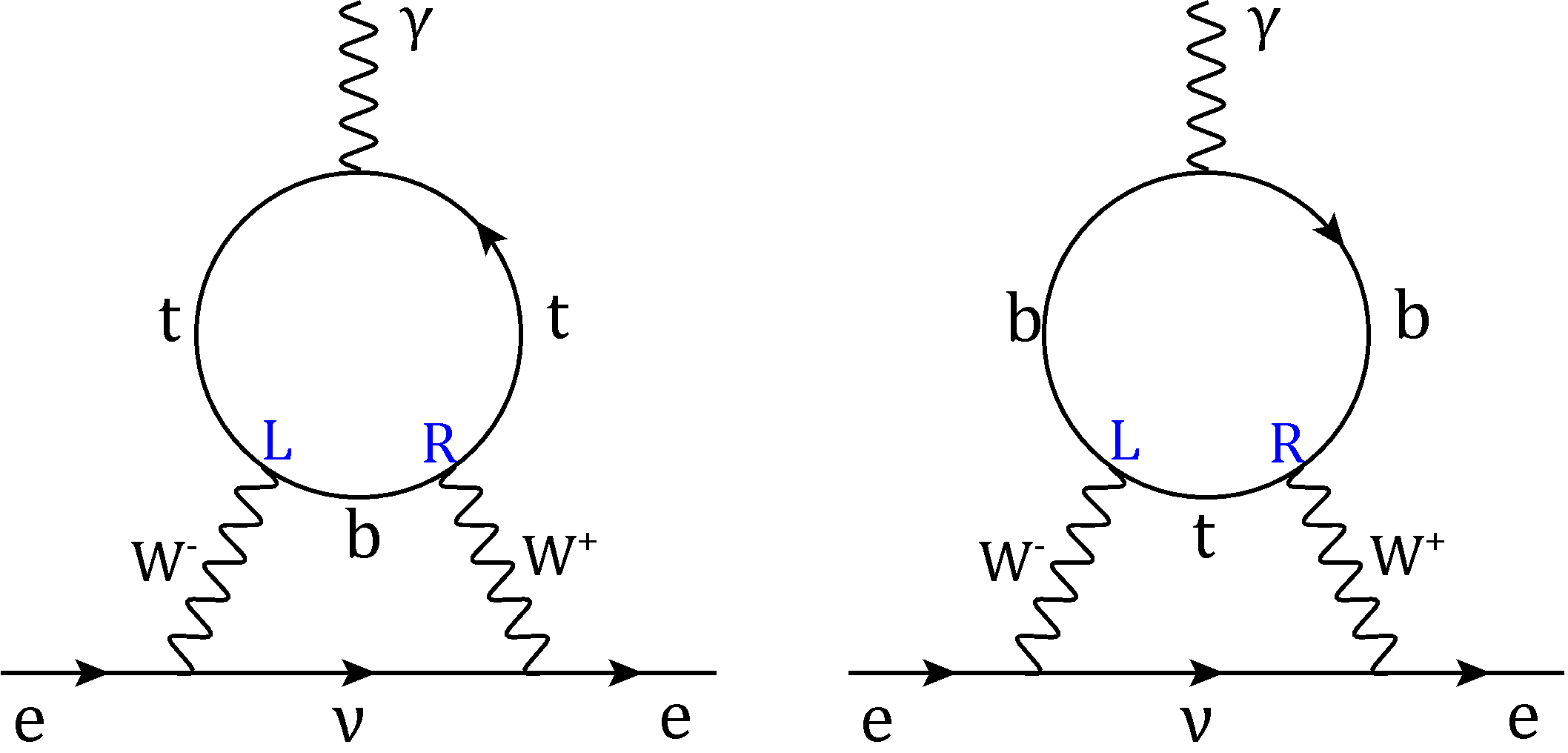}
\caption{Barr-Zee diagrams that contribute to electron EDM. The crossed diagrams with $L\leftrightarrow R$ are not shown.}
\label{fig:bz}
\end{figure}

Electric dipole moments are sensitive probes of new sources of CP violation. We first study the constraint from the  electron EDM. The interactions Eq.~(\ref{eq:Leff}) can contribute to the electron EDM
through the two-loop Barr-Zee type diagrams as shown in Fig.~\ref{fig:bz}.
This contribution has been calculated analytically in Ref.~\cite{AvilezLopez:2007wd},  
\begin{eqnarray}
\frac{d_e}{e} &=& - \frac{\alpha^2}{8\pi^2 \sin^4\theta_W M_W} z_ e z_t z_b {\rm Im}(a_L^* a_R) \nonumber \\
&\times&\frac{Q_b}{2} \int_0^1 d x_1 \int_0^{1-x_1} d x_2 
\left[ \frac{x(x-1)}{\left(g_b-x(1-x)\right)^2} \log \frac{g_b}{x(1-x)} - \frac{1}{g_b - x(1-x)} \right] - (b\leftrightarrow t) \,, \nonumber\\
\end{eqnarray}
where $x=x_1+x_2$, $z_a=m_a/M_W$, $(a=e,t,b)$, and $g_t=x(z_t^2-z_b^2)+z_b^2$, $g_b=x(z_b^2-z_t^2)+z_t^2$. Here $d_e$ is the coefficient of the effective EDM operator, 
\begin{eqnarray}
\mathcal{L}_{eff}\supset d_e \left(\frac{-i}{2} \right) \bar e \sigma_{\mu\nu} \gamma_5 e F^{\mu\nu}\,.
\end{eqnarray}
Numerically, we find,
\begin{eqnarray}
d_e\simeq -1.58\times10^{-27} {\rm Im}(a_L^* a_R)\ e\cdot{\rm cm} \,.
\end{eqnarray}
The current experimental upper limit on the electron EDM is $|d_e| < 8.7\times10^{-29} \ e\cdot{\rm cm}$ at 90\% CL from the ACME experiment in 2013~\cite{Baron:2013eja}. 
This translates into the upper bound,
\begin{eqnarray}
|{\rm Im}(a_R)|< 0.055 \,.
\end{eqnarray}
It turns out that the electron EDM constraint is weaker than the one from $B$ physics, as will be discussed in the next subsection, although the EDM constraint will become relevant if the current ACME limit is improved by only a factor of a few. In Fig.~\ref{fig:bp}, the horizontal magenta dotted line shows the future exclusion if the limit reaches 10 times the ACME-2013 limit.
%%
%The vectorlike quark mass is constrained by LHC to be $M\gtrsim800$\,GeV \cite{Aad:2015kqa}. 
%Therefore, for $|\lambda_t|\sim|\lambda_b|\sim1$, the ACME result still allows $\theta$ to be order 1.

Next, we consider the EDMs of the neutron, the proton and the mercury atom. These constraints usually involve large hadronic and nuclear physics uncertainties. However, given the future prospects of these experiments, they could become relevant. 
There are several contributions to these observables. 
The first includes light quark EDMs, from a similar diagram as Fig.~\ref{fig:bz}, with the replacement $(e, \nu) \to (u, d)$ or $(d, u)$. At $\mu=1\,$GeV,
\begin{eqnarray}
d_u (\mu)&\simeq& -2.3\times10^{-26} \eta_1
 {\rm Im}(a_L^* a_R)\ e\cdot{\rm cm} \,, \label{eq20} \\
d_d (\mu)&\simeq& -4.6\times10^{-26} \eta_1
 {\rm Im}(a_L^* a_R)\ e\cdot{\rm cm} \,. \label{eq21}
\end{eqnarray}
Here the renormalization group (RG) running effect from the  $M_W$ scale down to the GeV scale is taken into account, 
\begin{equation}
\eta_1 = \left[ \frac{\alpha_s(M_W)}{\alpha_s(m_b)} \right]^{\frac{16}{23}}
\left[ \frac{\alpha_s(m_b)}{\alpha_s(m_c)} \right]^{\frac{16}{25}}
\left[ \frac{\alpha_s(m_c)}{\alpha_s({1\, \rm GeV})} \right]^{\frac{16}{27}}
\simeq 0.417\, .
\end{equation} 
Hereafter we have used the NLO values of $\alpha_s$ at various scales in the following table.
\begin{table}[h]
\begin{tabular}{|c|c|c|c|}
\hline 
$\alpha_s(M_W)$ & $\alpha_s(m_b)$ & $\alpha_s(m_c)$  & $\alpha_s(1\,{\rm GeV})$ \\ 
\hline 
0.120808 & 0.218894 & 0.382156  & 0.455862 \\ 
\hline 
\end{tabular} 
\label{eq:VEVs}
\end{table}

The contribution of Eq.~(\ref{eq20}), (\ref{eq21}) to the neutron EDM, $d_n \sim - 0.35 d_u(\mu) + 1.4 d_d (\mu)$, is too small to yield a competitive constraint on ${\rm Im}(a_L^* a_R)$ in view of the current limit $|d_n| < 2.9\times10^{-26} \ e\cdot{\rm cm}$.

\begin{figure}[tb]
\centering
\includegraphics[width=0.33\textwidth]{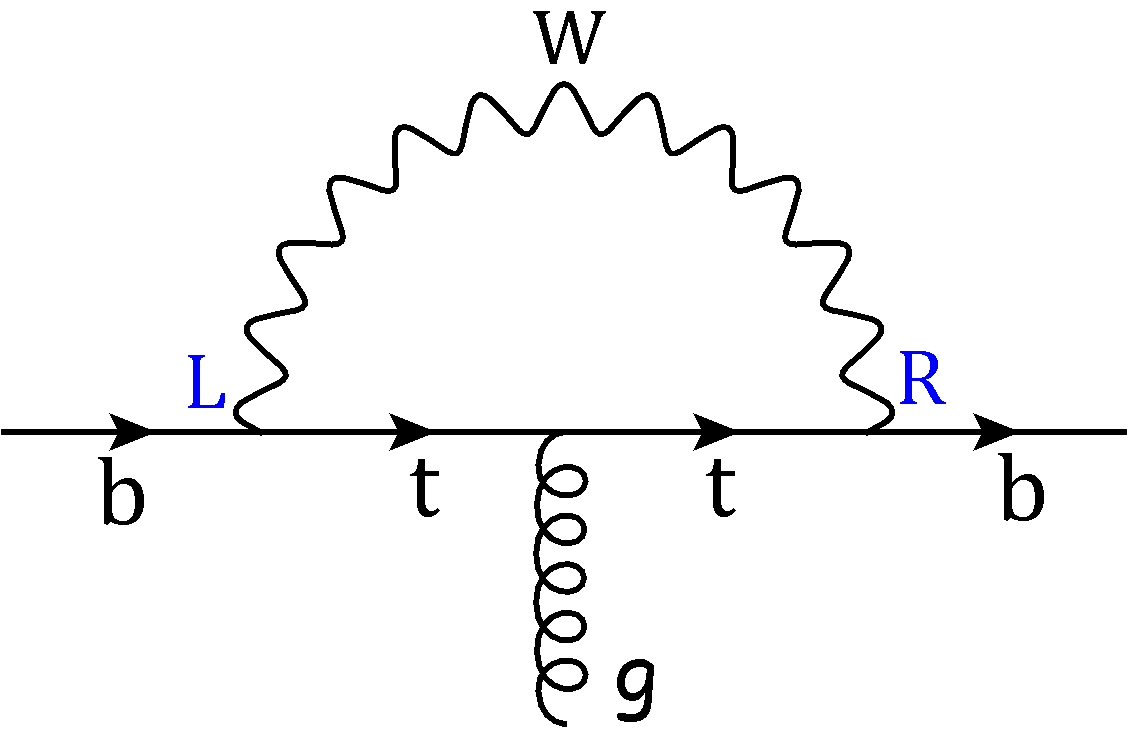}
\caption{One-loop contribution to the bottom quark chromo-EDM.}
\label{fig:1-loopCEDM}
\end{figure}

There is no light quark chromo-EDM at one loop level in the VLQ model. Instead, there is a contribution to the chromo-EDM of the bottom quark, shown in Fig.~\ref{fig:1-loopCEDM}\footnote{Unlike Ref.~\cite{AvilezLopez:2007wd}, we
find that diagrams similar to Fig.~\ref{fig:bz} but with photon lines replaced by gluon ones and leptons replaced by light quarks vanish and do not give rise to Chromo-EDMs.}. 
The bottom  quark chromo-EDM can contribute to the EDMs via matching to the three-gluon Weinberg operator at a low scale. 
The effective Lagrangian for the two operators takes the form~\cite{Engel:2013lsa},
\begin{eqnarray}
\mathcal{L}_{eff}\supset i \frac{\tilde \delta_b}{\Lambda^2} g_s m_b \bar b \sigma_{\mu\nu} \gamma_5 T^a b G^{a\mu\nu} + \frac{C_{\tilde G}}{2\Lambda^2} g_s f^{abc} \epsilon^{\mu\nu\rho\sigma} G_{\mu\lambda}^a G^{b\ \lambda}_\nu G^c_{\rho\sigma}\,.
\end{eqnarray}

The calculation of the coefficient of the chromo-EDM operator is similar to 
that in the left-right symmetric model~\cite{Xu:2009nt}. At the weak scale, 
\begin{eqnarray}
\frac{\tilde \delta_b(M_W)}{\Lambda^2} = - \frac{\sqrt{2} G_F}{8\pi^2} \frac{m_t}{m_b} {\rm Im}(a_L^* a_R) f(z_t) \,,
\end{eqnarray}
where $f(z_t)=\left[1-\frac{3}{4} z_t^2- \frac{1}{4} z_t^6+ \frac{3}{2} z_t^2\log z_t^2\right]/(1-z_t^2)^3 \simeq 0.35$.
Interestingly, there is an enhancement factor $(m_t/m_b)$~\cite{Chang:1990sfa}.
At the scale $m_b$, the matching condition is $C_{\tilde G}(m_b)=\frac{1}{12\pi}\alpha_s(m_b) \tilde \delta_b(m_b)$~\cite{Inoue:2014nva}.
Taking into account the RG running, the coefficient $C_{\tilde G}$ at $\mu=1\,$GeV is,
\begin{eqnarray}
\frac{C_{\tilde G} (\mu)}{\Lambda^2} &=& \frac{\alpha_s(m_b)}{12\pi}
\left[ \frac{\alpha_s(M_W)}{\alpha_s(m_b)} \right]^{\frac{14}{23}} 
\left[ \frac{\alpha_s(m_b)}{\alpha_s(m_c)} \right]^{\frac{29}{25}}
\left[ \frac{\alpha_s(m_c)}{\alpha_s(1\,\rm GeV)} \right]
\frac{\tilde \delta_b(M_W)}{\Lambda^2} \simeq -\frac{4.5\times10^{-9} {\rm Im}(a_L^* a_R)}{\rm GeV^2} \, \nonumber \\
\end{eqnarray}

The final contribution to the neutron EDM is dominated by the Weinberg operator~\cite{Engel:2013lsa, Inoue:2014nva},
\begin{eqnarray}
d_n &\simeq& (2\times 10^{-20} \ e\cdot {\rm cm}) \left( \frac{v^2}{\Lambda^2} \right) C_{\tilde G} (\mu) \simeq -5.5 \times 10^{-24} {\rm Im}(a_L^* a_R) \ e\cdot {\rm cm} \,,
\end{eqnarray}
where we have used the hadronic matrix element given in Ref.~\cite{Inoue:2014nva}.
The current limit on the neutron EDM, $|d_n| < 2.9\times10^{-26} \ e\cdot{\rm cm}$ at 95\% CL, translates into the upper bound,
\begin{eqnarray}
|{\rm Im}(a_L^* a_R)|< 0.5\times10^{-2} \,.
\end{eqnarray}
Because the three-gluon operator is an isospin singlet, the proton EDM in this model is equal to the neutron EDM. A possible future experiment measuring the proton EDM is expected to give a strong constraint~\cite{Kumar:2013qya, Yannis}.

The EDM of mercury $^{199}$Hg is also sensitive to the three-gluon operator, which contributes through the Schiff moment~\cite{Engel:2013lsa, Inoue:2014nva},
\begin{eqnarray}
d_{\rm Hg} = \kappa_S \frac{2 g_A m_N}{F_\pi} \left( a_0 \gamma^{\tilde G}_{(0)} + 
a_1 \gamma^{\tilde G}_{(1)} \right) \left( \frac{v^2}{\Lambda^2} \right) C_{\tilde G} \simeq 3.9\times10^{-27} {\rm Im}(a_L^* a_R) \ e\cdot {\rm cm} \,,
\end{eqnarray}
where we have used the conventions and values of parameters given in Ref.~\cite{Inoue:2014nva}.
The current limit on the mercury EDM, $|d_{\rm Hg}| < 3.1\times10^{-29} \ e\cdot{\rm cm}$, translates into the upper bound,
\begin{eqnarray}
|{\rm Im}(a_L^* a_R)|< 0.8\times10^{-2} \,.
\end{eqnarray}
%We want to emphasize that the above current limits are obtained using the central values of the relevant nuclear and hadronic matrix elements, which are known to have large uncertainties. 

\subsection{B physics}
There are strong constraints on the parameter ${\rm Im}(a_L^*a_R)$ from the $b \to s \gamma$ channel, both from the total rate and the CP asymmetry. 
The effective Lagrangian relevant for this process takes the form,
\begin{eqnarray}
\mathcal{L}_{eff}^{(b\to s\gamma)} &=& c_7 {e m_b \over 16 \pi^2} {\bar s}_L \sigma^{\mu\nu} b_R F_{\mu\nu} + c_8 {g_s m_b \over 16 \pi^2} {\bar s}_L \sigma^{\mu\nu} T^a b_R G^a_{\mu\nu} \,.
\end{eqnarray}
The contribution of new right-handed current interaction to the Wilson coefficients at the scale $M_W$ are~\cite{Grzadkowski:2008mf},
\begin{eqnarray}
\Delta c_7(M_W) = %a_L c_7^{\rm SM}(\Lambda) + 
a_R \frac{m_t}{m_b} f_7 (m_t^2/M_W^2) \,, \nonumber \\
\Delta c_8(M_W) = %a_L c_8^{\rm SM}(\Lambda) + 
a_R \frac{m_t}{m_b} f_8 (m_t^2/M_W^2) \,,
\end{eqnarray}
where the form factors are,
\begin{eqnarray}
f_7 (x) &=& \frac{-3x^2+2x}{2(x-1)^3} \log x + \frac{-5 x^2+31 x -20}{12(x-1)^2} \,, \nonumber \\
f_8 (x) &=& \frac{3x}{2(x-1)^3} \log x - \frac{x^2+ x +4}{4(x-1)^2} \,.
\end{eqnarray}
When we take into account  the 1-loop RG running corrections from $M_W$ to $\mu=m_b$, the effective coefficients are~\cite{Kagan:1998bh},
\begin{eqnarray}
c_7(\mu) &=& a_L c_7^{\rm SM}(\mu) + \eta_b^{16/23} \Delta c_7(M_W) + \frac{8}{3} (\eta_b^{14/23} - \eta_b^{16/23}) \Delta c_8(M_W) \,, \nonumber \\
c_8(\mu) &=& a_L c_8^{\rm SM}(\mu) + \eta_b^{14/23} \Delta c_8(M_W) \,,
\end{eqnarray}
with $c_7^{\rm SM}(\mu)=-0.31$, $c_8^{\rm SM}(\mu)=-0.15$ and $\eta_b=\alpha_s(M_W)/\alpha_s(m_b)\simeq0.55$.

The $B\to X_s \gamma$ decay rate in the VLQ model is then given by
\begin{eqnarray}
\mathcal{B}(B\to X_s \gamma) = \mathcal{B}(B\to X_s \gamma)_{\rm SM} \left|\frac{c_7}{c_7^{\rm SM}}\right|^2 \,.
\end{eqnarray}
The SM prediction has a central value $\mathcal{B}(B\to X_s \gamma)_{\rm SM}=3.15\times10^{-4}$. The world average of the  measurements is
$\mathcal{B}(B\to X_s \gamma)=(3.55\pm0.24\pm0.09) \times 10^{-4}$~\cite{Asner:2010qj}.

%%%%%%%%%%%%%%%%%%%%%%%%%%%%%%%%%
\begin{figure}[t]
\includegraphics[width=0.75\textwidth]{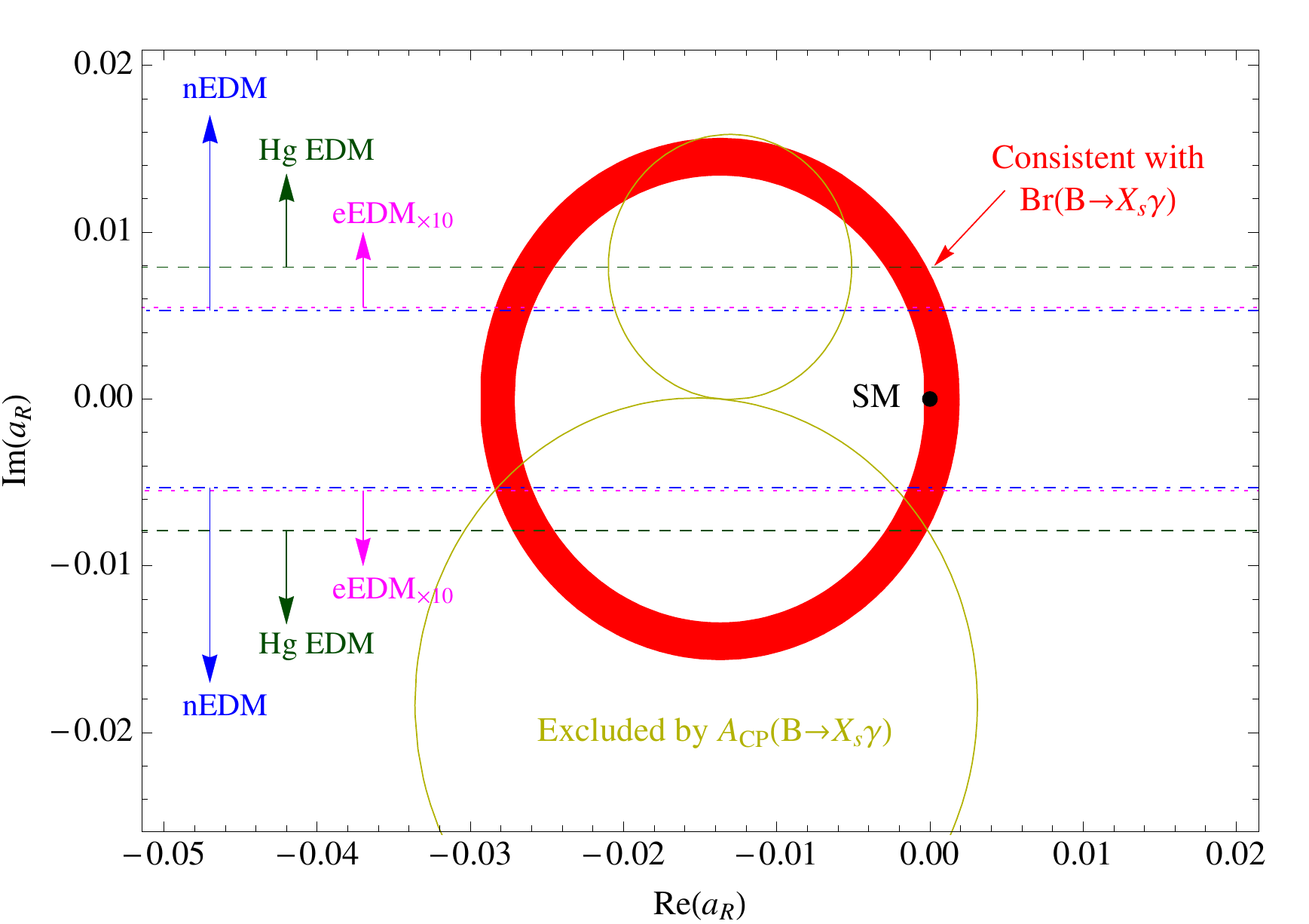}
\caption{Constraints on the Re($a_R$) and Im($a_R$) parameter space from EDMs and 
the $B\to X_s \gamma$ decay.
The red shaded region is consistent with the branching ratio $\mathcal{B}(B\to X_s \gamma)$. 
The regions inside the solid yellow circles are excluded by the direct CP asymmetry, $A_{CP}$,
 in $B\to X_s \gamma$ decay.
The EDM constraints are shown by the blue dot-dashed lines (current neutron EDM, central value),
green dashed lines (current mercury EDM) and magenta dotted lines (future electron EDM with 10 times better limit than the current one from ACME). The exclusion regions are in the direction of arrows.
The neutron and mercury EDM typically involve large
nuclear/hadronic uncertainties, and here the lines only show the central value of their bounds.
}
\label{fig:bp}
\end{figure}
%%%%%%%%%%%%%%%%%%%%%%%%%%%%%%%%

In order to present this result as a limit, we note the fact that in the VLQ model
the contribution to $a_R$ dominates over the deviation $a_L-1$. 
Using results from the EDM discussions that ${\rm Im}(a_L^* a_R)$ is already constrained to be less than $\sim 10^{-2}$, we conclude that 
\begin{eqnarray}
{\rm Im}(a_L^* a_R) \simeq {\rm Im}(a_R) \,.
\end{eqnarray}
In this case, the constraint from the $B\to X_s \gamma$ decay rate measurement is presented in the parameter space of ${\rm Re}(a_R)$ versus ${\rm Im}(a_R)$, as shown by the red shaded region in Fig.~\ref{fig:bp}.

The direct CP asymmetry in $B\to X_s \gamma$ decay rate is~\cite{Benzke:2010tq}
\begin{eqnarray}
A_{CP} = \alpha_s(m_b) \Biggl\{ \frac{40}{81} {\rm Im} \left(\frac{c_2}{c_7}\right) - \frac{4}{9} {\rm Im} \left(\frac{c_8}{c_7}\right) - \frac{40 \Lambda_c}{9 m_b} {\rm Im} \left[ (1+\epsilon_s) \frac{c_2}{c_7} \right] \Biggr\} \,,
\end{eqnarray}
where $c_2=1.11$, $\epsilon_s = -0.007+0.018i$, 
$\Lambda_c=0.38\,$GeV, and $\alpha_s(m_b)=\alpha_s(M_W)/\eta=0.21$. The most stringent experimental measurement is from BaBar, $A_{CP}=(1.7\pm1.9\pm1.0)\%$~\cite{Lees:2014uoa}.
Again, we show this as a constraint in the ${\rm Re}(a_R)$ versus ${\rm Im}(a_R)$ plane in Fig.~\ref{fig:bp}. The regions inside the yellow circles are excluded.

Summarizing the EDM and the $B$ physics constraints, we conclude that in the doublet VLQ model it is still possible to have ${\rm Im}(a_R)$ as large as order 0.01.
From Eq.~(\ref{eq:a3w}), this implies the CP violating $hWW$ coupling $a_3^W$ is currently constrained to be at most $10^{-5}$. 
The next generation EDM search is expected to further narrow down the allowed window of ${\rm Im}(a_R)$. In the case of discovery, this would trigger an exciting interplay between the studies of CP violation in a future $B$ factory and a future Higgs factory.

\section{Conclusion}
\label{sec:conc}
In this work, we have studied the possibility of introducing CP violating interactions to the 125 GeV Higgs boson by extending the fermion sector of the SM with vectorlike quarks. We examined the simplest class of models where VLQs arise from a single representation under the SM gauge group. There are  7 possible representations where the VLQs have Yukawa interactions with the SM third generation quarks and Higgs doublets. 
The new complex Yukawa couplings could accommodate new sources of CP violation.
Among them, we find that an irreducible CP phase shows up only for one representation of VLQ, which is a doublet under $SU(2)_L$ and carries hypercharge $1/3$.
For the other representations all the phases can be rotated away and are unphysical.

We study the CP violating phenomenology in the doublet VLQ model.
Since the VLQs are already constrained to be heavier than 800 GeV by the LHC, we integrate them out and study  the effective theory, where CP violation manifests itself through a new right-handed charged current mediated by the SM $W$-boson.
We have calculated the CP violating Higgs interactions with SM gauge bosons, generated at one loop level involving both top and bottom quarks. This corresponds to a dimension 8 operator in the heavy top/bottom quark limit. 
As a consequence, only the $hWW$ coupling is CP violating, while the $hZZ$, $h\gamma \gamma$, $hZ\gamma$ couplings are essentially CP conserving at this order. 
The strength of the CP violating $hWW$ coupling is proportional to
the quantity ${\rm Im}(a_L^* a_R)$, where $a_{L,R}$ are the coefficients of left- and right-handed current $Wtb$ interactions, respectively.
At low energy, we find the most relevant constraints on ${\rm Im}(a_L^* a_R)$ come from the electric dipole moments and  the $b\to s\gamma$ decay rate and CP asymmetry, which are in complimentary to each other. The current experimental constraints require  ${\rm Im}(a_L^* a_R) \lesssim0.01$. They in turn imply that the coefficient of the CP violating $hWW$ interaction, $a_3^W$, 
cannot be larger than of order $10^{-5}$, and, as we stress again, only in the $hWW$ channel.
We expect exciting interplays of various experimental searches in the future to probe and distinguish new sources of CP violation near the electroweak scale.

%\appendix
%\numberwithin{equation}{section}
%\section{}
%\label{appa}

\acknowledgements 
We thank JiJi Fan, Enrico Lunghi, and Miha Nemevsek for useful discussions.
The work of C.-Y. Chen and S. Dawson is supported by the U.S. Department of Energy under grant No. DE-AC02-98CH10886 and contract DE-AC02-76SF00515. This work of Y. Zhang is supported by the Gordon and Betty Moore Foundation through Grant \#776 to the Caltech Moore Center for Theoretical Cosmology and Physics, and by the DOE Grant DE-FG02-92ER40701, and also by a DOE Early Career Award under Grant No. DE-SC0010255.

%\begin{thebibliography}{99}
%\cite{Aguilar-Saavedra:2013qpa}
%\bibitem{Aguilar-Saavedra:2013qpa} 
 % J.~A.~Aguilar-Saavedra, R.~Benbrik, S.~Heinemeyer and M.~P?rez-Victoria,
  %``Handbook of vectorlike quarks: Mixing and single production,''
 % Phys.\ Rev.\ D {\bf 88}, no. 9, 094010 (2013)
  %[arXiv:1306.0572 [hep-ph]].
  %%CITATION = ARXIV:1306.0572;%%
  %73 citations counted in INSPIRE as of 25 juin 2015

%\end{thebibliography}
%\newpage

\bibliographystyle{unsrt}

\bibliography{vlcpv.bib}

\end{document}